\documentclass[twocolumn,showpacs,preprintnumbers,amsmath,amssymb,prd,aps]{revtex4-1}

\usepackage{graphicx}

\begin{document}

\title{What have we learned about the sources of ultrahigh-energy cosmic rays via neutrino astronomy?}


\author{Shigeru Yoshida}

\affiliation{Department of Physics and Institute for Global Prominent Research, 
Chiba University, Chiba 263-8522, Japan}

\begin{abstract}
Observations of TeV--PeV-energy cosmic neutrinos by the IceCube observatory have suggested that
extragalactic cosmic-ray sources should have an optical depth greater than $\sim$0.01
and contribute to more than 10\% of the observed bulk of cosmic rays at 10 PeV. 
If the spectrum of cosmic rays from these extragalactic sources extends well beyond 1 EeV, 
the neutrino flux indicates that extragalactic cosmic-ray protons are dominant 
in the observed total cosmic-ray flux at 1 EeV.
Among known powerful astronomical objects,
including gamma-ray bursters (GRBs), only flat-spectrum radio quasars could (barely) satisfy these conditions. 
On the other hand, the null detection of neutrinos with energies
well beyond PeV has excluded the possibility that radio-loud active galactic nuclei (AGNs) 
and/or GRBs, the popular source candidates discussed in the literature, 
are the origins of the highest-energy cosmic rays
($\sim100 {\rm EeV}$) if they are composed mainly of protons.
Their origins must be objects that have evolved on time scales comparable to or slower than the star formation rate.
These considerations indicate that none of the known
extragalactic astronomical objects can be simultaneously a source of both PeV- and trans-EeV-energy
cosmic rays. As a result of the stringent limits on EeV-energy neutrino fluxes, 
a significant part of the parameter space for the AGN
and new-born pulsar models is starting to seem unfavorable, even for scenarios of mixed and heavy cosmic-ray compositions
at the highest energies.
\end{abstract}

\maketitle


\section{Introduction}
\label{sec:introduction}

High-energy neutrino astronomy has finally begun. 
The detection of PeV-energy neutrinos~\cite{icecubePeV2013} and the follow-up
analyses~\cite{icecubeHESE2013} by the IceCube Collaboration revealed the existence
of astrophysical ``on-source'' neutrinos at energies ranging from TeV to PeV.
These neutrinos are expected to be produced by the interactions
of ultrahigh-energy cosmic-ray (UHECR) protons via $pp$ collisions
or $\gamma p$ collisions. The bulk intensity of these neutrinos,
$E_\nu^2 \phi_{\nu_e+\nu_\mu+\nu_\tau}\simeq 
3.6\times 10^{-8} {\rm GeV} {\rm cm^{-2}} \sec^{-1} {\rm sr^{-1}}$,
provides an important clue to understanding the general characteristics
of UHECR sources through the connection between the observed cosmic-ray
and neutrino intensities. 

In the even higher-energy region from EeV to 100 EeV (EeV $=10^9 {\rm GeV}$),
the highest-energy cosmic-ray (HECR) protons generate EeV-energy neutrinos
via interactions with cosmic microwave background (CMB) photons~\cite{GZK} and extragalactic background light (EBL)
during their propagation in intergalactic space. The intensity of these
``GZK cosmogenic'' neutrinos~\cite{BZ} averaged over the sky is a consequence
of the integral of the HECR emission over cosmic time, as neutrinos are
strongly penetrating particles that can travel cosmological
distances. It is, therefore, an observational probe to trace
the HECR source evolution. In particular, the cosmogenic neutrino intensity
from 100 PeV to 10 EeV is highly sensitive to the evolution of the HECR emission rate
and less dependent on other uncertain factors such as the highest energy
of accelerated cosmic rays at their sources. As this energy range coincides with
the central region covered by the IceCube ultrahigh-energy neutrino searches,
the flux sensitivity achieved by IceCube has started to constrain a sizable
parameter space of HECR source evolution, revealing the general
characteristics of UHECR and HECR sources independent of the cosmic-ray acceleration model.

In this article, we review the new knowledge of UHECR/HECR sources
provided by neutrino observations by IceCube. 
The standard $\Lambda$CDM cosmology with $H_0 = 73.5$ km s$^{-1}$ Mpc$^{-1}$, $\Omega_{\rm M} = 0.3$, 
and $\Omega_{\Lambda}=0.7$ is assumed throughout this article.

\section{The cosmic neutrino spectrum: Overview}
\label{sec:overview}

\begin{figure}
\includegraphics[width=0.45\textwidth]{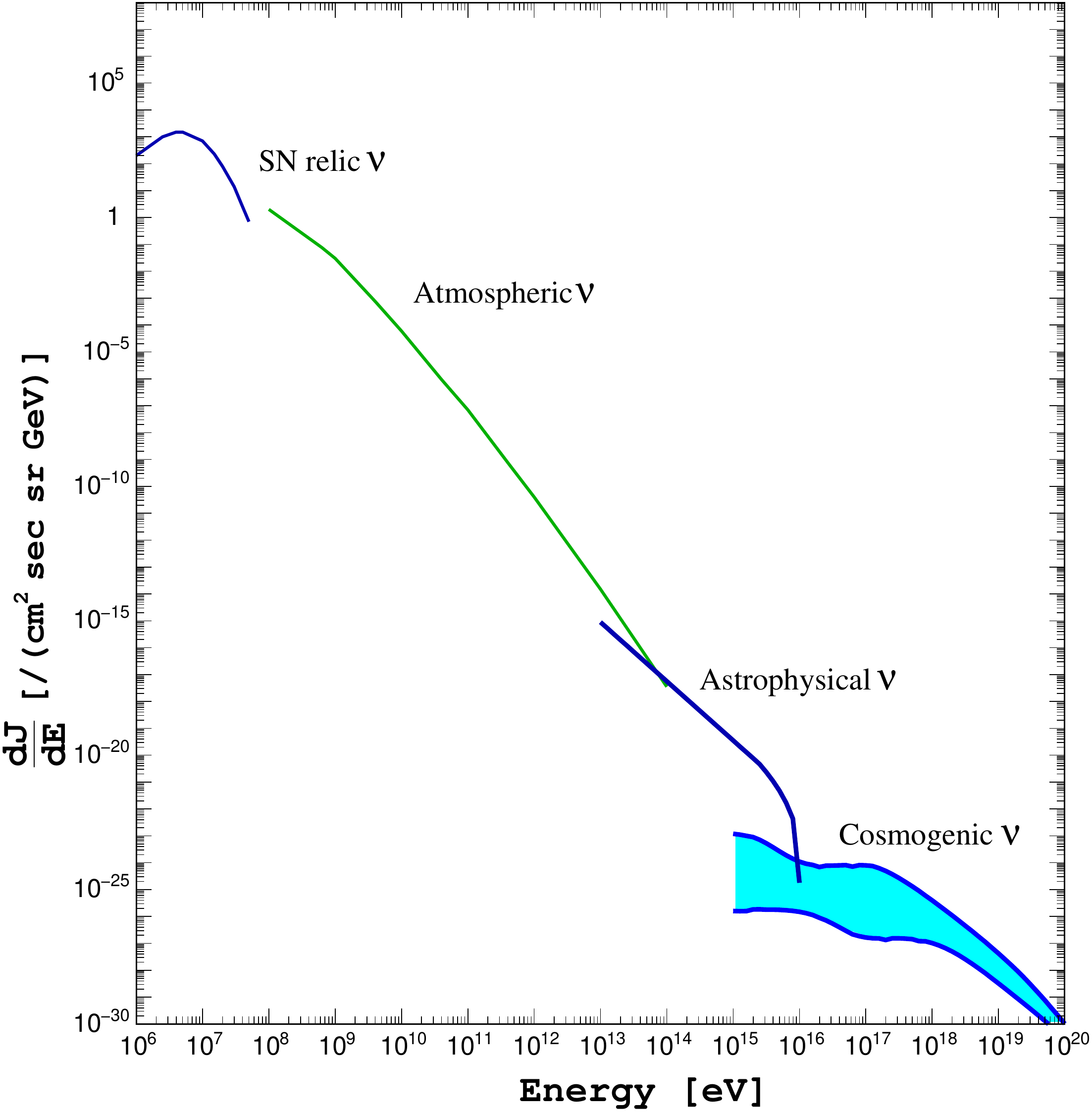}
\caption{Illustrative display of the differential fluxes of diffuse neutrinos having various origins. 
Supernova relic neutrinos are expected to appear in the MeV sky. Atmospheric
neutrinos originating in extensive cosmic-ray airshowers dominate the background
for high-energy cosmic neutrino detection. Astrophysical neutrinos produced
{\it in situ} at cosmic-ray sources emerge at TeV and PeV energies, which have now been
detected by IceCube. GZK cosmogenic neutrinos, whose flux has yet to be measured, fill the highest-energy universe. Shown here is the possible range of fluxes
from various source evolution models. The proton-dominated HECR composition is assumed.}
\label{fig:neut_fluxes}
\end{figure}

Figure~\ref{fig:neut_fluxes} displays the spectrum of neutrinos coming from
all over the sky, {\it i.e.}, diffuse neutrino fluxes. The massive background
of atmospheric neutrinos ranging in energy over many orders of magnitude
had masked the astrophysical neutrinos until the IceCube observatory finally
revealed their existence. The spectrum shown here, taken from a model~\cite{YoshidaTakami2014},
has a cutoff feature at $\sim{\rm PeV}$, but it is not clear yet whether the IceCube neutrino
fluxes have a spectral cutoff. It is not even obvious that the spectrum can be well
described by a single power law formula. An analysis with enhanced sensitivity
in the 10 TeV region seems to exhibit a trend toward a softer spectrum~\cite{MESE} than
the up-going diffuse muon neutrino analysis, which is sensitive
at energies above $\sim 100\ {\rm TeV}$~\cite{diffuse_nu}.
On-going efforts in the IceCube Collaboration will ultimately resolve these 
issues. Nevertheless, the intensity, 
$E_\nu^2 \phi_{\nu_e+\nu_\mu+\nu_\tau}\simeq 
3.6\times 10^{-8} {\rm GeV} {\rm cm^{-2}} \sec^{-1} {\rm sr^{-1}}$, 
has been well determined within a factor of two, and the implications
for the origin of UHECRs based on the intensity are not affected by these
details of the spectral structure. If the neutrino emitters are also
sources of the cosmic rays we are observing (which is very likely but not an undeniable assumption),
we can associate the neutrino flux with their parent cosmic-ray proton flux, 
and its comparison to the {\it observed} cosmic-ray spectrum places
some constraints on the source characteristics.

The GZK cosmogenic neutrinos are expected to emerge in the 100 PeV--EeV sky.
Their intensity at the highest-energy end ($\sim50-100\ {\rm EeV}$) depends mainly on
the maximal accelerated energy of cosmic rays at their sources
and is not relevant to the ultrahigh-energy neutrino search by IceCube,
as it is most sensitive at energies below 10 EeV.
The intensity at the lowest-energy tail ($\sim10-100\ {\rm PeV}$)
is determined by the EBL density
and its evolution, which is EBL-model-dependent and could vary
the flux by a factor of $\sim5$ at $\sim10\ {\rm PeV}$.
The EeV-energy intensity is decided primarily by the HECR source evolution
in redshift space.
The integral intensity of cosmogenic neutrinos above 100 PeV ranges from $10^{-17}$ to 
$\sim3\times 10^{-16} {\rm cm^{-2}} \sec^{-1} {\rm sr^{-1}}$
depending on these factors. The IceCube detection exposure
for UHE neutrinos has now reached $\sim3\times 10^{16} {\rm cm^{2}} \sec {\rm sr}$,
and one can see that the IceCube sensitivity enables access to
a significant parameter space of the cosmogenic neutrino production models.

\section{The constraints on PeV- and EeV-energy UHECR sources}
\label{sec:PeV-EeV}

The flux of astrophysical neutrinos produced by UHECR protons at
their sources is related to their parent cosmic-ray intensity
via the proton-to-neutrino conversion efficiency. The efficiency is usually
parameterized in the form of the optical depth of the proton interactions.
For neutrinos produced through photomeson production ($\gamma p$),
their diffuse flux integrating emitted neutrinos over all the sources
of UHECR protons with a spectrum in the source
frame, $\sim \kappa_{\rm CR}(E_{\rm CR}/E_0)^{-\alpha}$ (where $E_0$ is the reference energy,
which is conveniently set to $\sim10$ PeV), is described as~\cite{YoshidaTakami2014}
\begin{widetext}
\begin{eqnarray}
\phi_{\nu_e + \nu_\mu + \nu_\tau}(E_\nu) &\simeq& 
\frac{2 n_0 \kappa_{\rm CR}}{\alpha^2} \frac{c}{H_0} 
\frac{s_{\rm R}}{\sqrt{(s_{\rm R} + m_{\pi}^2 - m_p^2)^2 - 4 s_{\rm R} m_{\pi}^2}} 
\frac{3}{1 - r_{\pi}} 
\frac{(1 - e^{-\tau_0})}{2 (m - \alpha) - 1} \Omega_{\rm M}^{- \frac{m - \alpha + 1}{3}} \nonumber \\
&& ~~~~~~~~~~~~~~~~~~~~~~~~~~~~~~~ \times 
\left[ \left\{ \Omega_{\rm M} (1 + z_{\rm max})^3 + \Omega_{\Lambda} \right\}^{\frac{m - \alpha}{3} - \frac{1}{6}} - 1 \right] 
\left( \frac{E_{\nu}}{E_0 x_{\rm R}^+ (1 - r_{\pi})} \right)^{-\alpha}. 
\label{eq:onsource_simple}
\end{eqnarray}
\end{widetext}
Here $n_0$ is the source number density at the present epoch,
and the source evolution is parameterized as $\psi(z)=(1+z)^m$
extending to the maximal redshift $z_{\rm max}$
such that the parameter $m$ represents the scale of the cosmological
evolution often used in the literature. Further, $s_R\ (\simeq 1.5 {\rm GeV^2})$ 
is the squared collision energy at the $\Delta$ resonance of photopion production,
$r_{\pi} \equiv m_{\mu}^2 / m_{\pi}^2 \simeq 0.57$ is the muon-to-pion mass-squared ratio,
$m_p$ is the proton mass, and $x_{\rm R}^+\ (\simeq 0.36)$ 
is the kinematically maximal bound 
of the relative energy of emitted pions normalized by the parent cosmic-ray energy.
$\tau_0$, the optical depth of $\gamma p$ interactions at the reference energy $E_0$,
links the neutrino flux to the parent cosmic-ray intensity determined by $\kappa_{\rm CR}$.

The cosmic-ray flux integrating a UHECR spectrum over all
the sources in the redshift space, which corresponds to the UHECR spectrum
we {\it observe}, is given by
\begin{eqnarray}
\phi_{\rm CR}(E_{\rm CR}) &=& 
n_0 c \kappa_{\rm CR} \int_0^{z_{\rm max}} dz \nonumber \\
&& (1 + z)^{1 - \alpha} \psi(z) \left| \frac{dt}{dz} \right| 
e^{- \tau_0} 
\left( \frac{E_{\rm CR}}{E_0} \right)^{-\alpha},\nonumber\\
&&
\end{eqnarray}
neglecting intergalactic magnetic fields and the energy loss
in the CMB field during UHECR propagation.
Introducing some analytical approximations leads to the following simple formula:
\begin{eqnarray}
\phi_{\rm CR}(E_{\rm CR}) &\simeq& 
2n_0 \kappa_{\rm CR} \frac{H_0}{c} e^{-\tau_0}\left( \frac{E_{\rm CR}}{E_0} \right)^{-\alpha}\nonumber\\
&& \frac{1}{2(m-\alpha)-1} \Omega_{\rm M}^{- \frac{m-\alpha+1}{3}}\nonumber\\
&&\left[ \left\{ \Omega_{\rm M} (1 + z_{\rm max})^3 + \Omega_{\Lambda} \right\}^{\frac{m-\alpha}{3} - \frac{1}{6}} - 1 \right].\nonumber\\
&& 
\label{eq:UHECR_flux_approx}
\end{eqnarray}

Comparing this formula to Equation~(\ref{eq:onsource_simple}),
one can find that the source evolution effect represented by the evolution
parameter $m$ is canceled in the ratio of the neutrino flux
to the parent UHECR flux. This is because both the secondary produced neutrinos
and emitted UHECRs originate in the same sources with the same evolution history.
Consequently, the optical depth $\tau_0$ is a deciding factor
in the relation between these two fluxes. The TeV--PeV neutrino observation
by IceCube that determines the neutrino flux $\phi_\nu$ thus
associates the UHECR source optical depth with the UHECR flux.

\begin{figure}[tb]
  \includegraphics[width=0.4\textwidth]{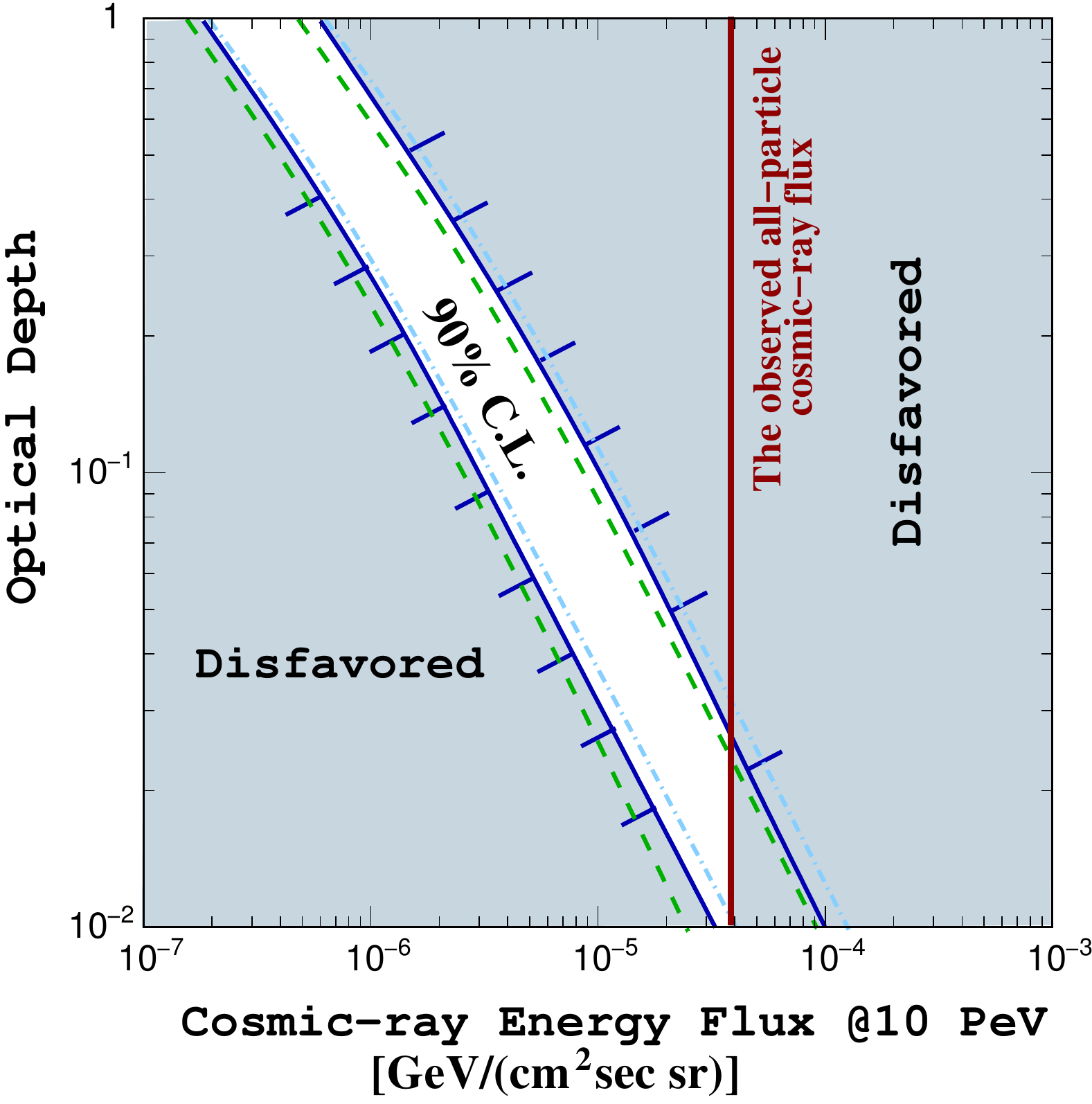}
  \caption{Constraints on the optical depth of UHECR sources 
for PeV-energy neutrino production and the energy flux of 
extragalactic UHECRs, $E_{\rm CR}^2\phi_{\rm CR}$, at an energy of 10 PeV~\cite{YoshidaTakami2014}. 
The regions between the two blue solid curves ($\alpha=2.5$), 
green dashed curves ($\alpha=2.7$), and light blue dot-dashed curves ($\alpha=2.3$) 
are allowed by the present IceCube observations~\cite{icecubePeV2013,icecubeHESE2013}.
The unshaded region highlights the allowed region for $\alpha=2.5$ 
taking into account the observed intensity of UHECRs measured by the IceTop experiment~\cite{icetop2013}.}
\label{fig:constraints_10PeV} 
\end{figure}

\begin{figure}
  \includegraphics[width=0.4\textwidth]{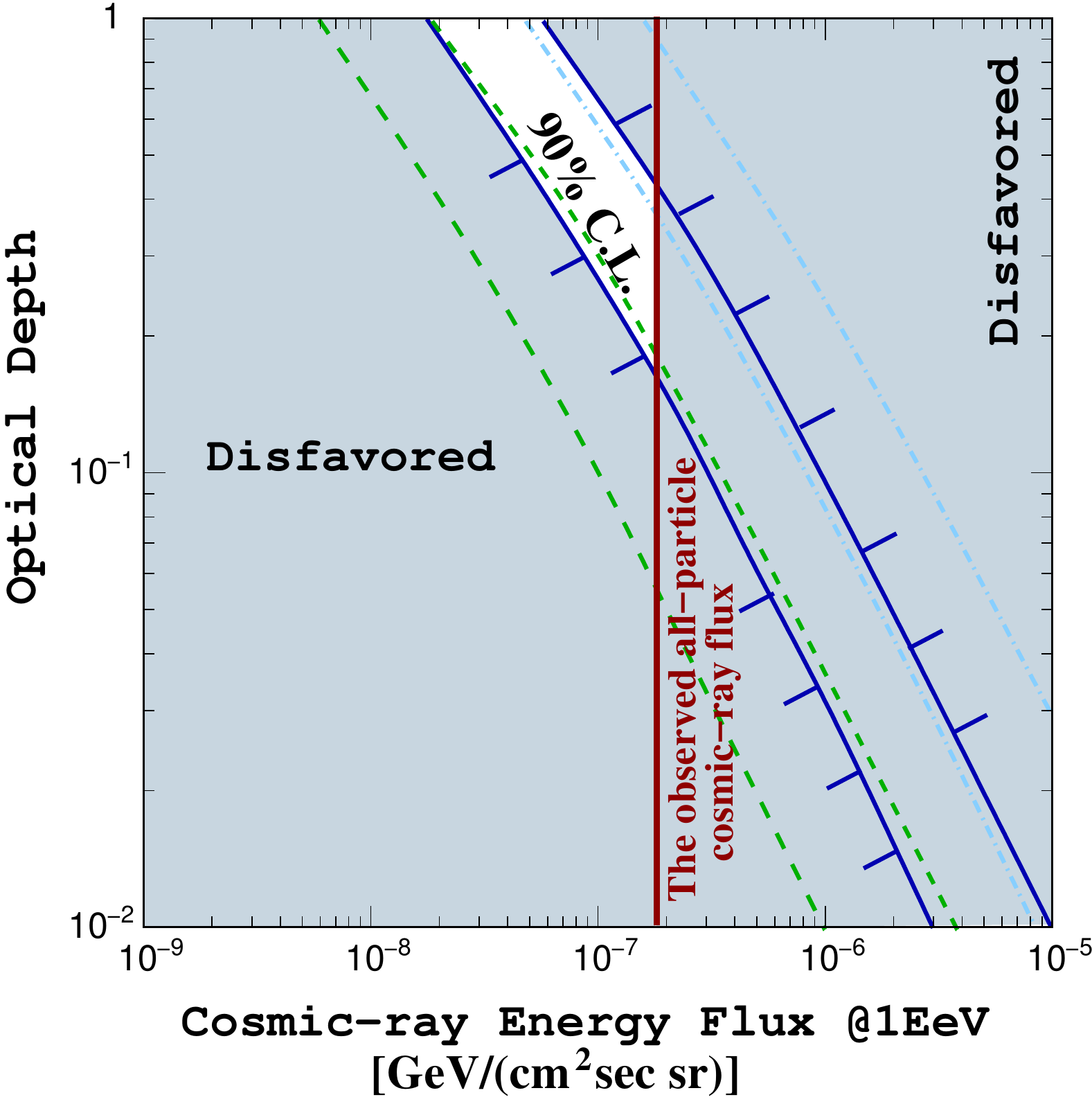}
  \caption{Same as Figure~\ref{fig:constraints_10PeV}, but constraints against the cosmic-ray flux
    at an energy of 1 EeV, showing the constraints when the UHECR spectrum from PeV neutrino
    sources extends to higher energies.}
\label{fig:constraints_1EeV} 
\end{figure}

Figure~\ref{fig:constraints_10PeV}
displays the relations between the optical depth and the UHECR flux for several values of $\alpha$,
all of which are consistent with the IceCube observation
at the present statistics~\cite{icecubeHESE2013}. Star-formation-like source evolution
is assumed, but the other assumptions regarding the evolution 
would not change the main results, as explained above.
The smaller optical depth, implying a lower neutrino production efficiency, would require
more UHECR protons to be compatible with the neutrino intensity measured by IceCube.
The optical depth of the UHECR sources must be larger than 0.01,
as the parent UHECR flux would exceed the observed cosmic-ray flux otherwise.
The proton flux from the neutrino sources contributes more than at least a few percent
of all the UHECRs in the 10-PeV energy range. The magnetic horizon effect
would not change these constraints unless the sources are very rare, for example, 
if their number density is much smaller than $\sim10^{-6} {\rm Mpc}^{-3}$~\cite{YoshidaTakami2014}.
Note that the lower bound of the source density set by the small-scale UHECR anisotropy study
conducted by the Auger observatory~\cite{auger_density} is $6\times 10^{-6} {\rm Mpc}^{-3}$.

Gamma-ray bursters (GRBs) are strong candidates for UHECR acceleration sites
and therefore high-energy neutrino production sites. 
Internal shocks are the most popular sites to produce
high-energy neutrinos. An optical depth of $0.1-10^{-2}$ can be achieved,
depending on the dissipation radius, which satisfies the optical depth condition
shown in Figure~\ref{fig:constraints_10PeV}.
However, their energetics may be problematic.
The typical gamma-ray energy output of a
regular GRB is $10^{52}$ erg in gamma rays, and the local
occurrence rate of long-duration GRBs is $\sim1 {\rm Gpc^{-3}}\ {\rm yr^{-1}}$.
These data indicate that the luminosity of local cosmic rays generated from GRBs is
$10^{44}(\eta_p/10) {\rm erg}\ {\rm Mpc^{-3}}\ {\rm yr^{-1}}$,
where $\eta_p$ is the ratio of
the UHECR output and $\gamma$-ray output, which is known as the
baryon loading factor. This luminosity is 2 orders
of magnitude smaller than that of UHECRs at 10 PeV and
thus is too low to meet the requirement shown in Figure~\ref{fig:constraints_10PeV}
that the UHECR flux from the sources must account for $\geq0.1$ of the total UHECR flux.
We conclude that 
GRBs are unlikely to be major sources of both PeV-energy UHECRs and neutrinos.

Among known astronomical objects,
only flat-spectrum radio quasars (FSRQs)
can realize a large $\gamma p$ optical depth ($\geq0.01$)
and large energetics. The typical $\gamma$-ray luminosity density
of FSRQs is $\sim10^{46} {\rm erg}\ {\rm Mpc^{-3}}\ {\rm yr^{-1}}$
in our local universe. This is comparable to the local density
of UHECRs at $\sim$10 PeV.

Figure~\ref{fig:constraints_1EeV} shows the constraints on the $\gamma p$ optical depth
and UHECR flux when the energy spectrum of UHECR protons emitted from the neutrino sources
extends to much higher energies than the observed neutrino energies.
The allowed regions in the parameter space become much smaller than those
for the constraints for $E_0 = 10$ PeV because the spectrum of the observed cosmic rays
is steeper than the observed neutrino spectrum.
Note that the optical depth constrained here is not at the EeV level, but at the PeV level,
because PeV-energy protons are responsible for the neutrinos detected by IceCube.
The constraints suggest that the optical depth of protons for PeV-energy
neutrinos is rather high, $\tau_0\geq 0.2$, and also that a major
fraction of UHECRs in the EeV region is extragalactic
protons. This supports the ``dip'' transition model~\cite{dip_model} of UHECR protons, 
where the ankle structure of the cosmic-ray spectrum, which
appears at 3 to 10 EeV, is caused by the energy loss of
extragalactic UHECR protons by Bethe--Heitler pair production with
CMB photons. This model predicts a high GZK cosmogenic neutrino flux 
at 10--100 PeV.
However, as we see in the next section,
the null detection of cosmogenic neutrino candidates in the IceCube seven year dataset
excludes the dip transition scenario if HECRs are proton-dominated.
This suggests that the source of neutrinos
seen by IceCube is not the main source of cosmic rays at EeV energies or higher.
No known class of astronomical objects can 
meet the stringent requirement of the $\gamma p$ optical depth, $\tau_0\geq 0.1$,
and the UHECR energetics. Only bright FSRQs such as those with 
$L_\gamma\sim 10^{50} {\rm erg}\ \sec^{-1}$ {\it could} realize
$\tau_0\sim 0.1$, but such FSRQs are too rare to energetically reproduce
the observed UHECR flux at 1 EeV.

According to these considerations, none of the known extragalactic objects
can be found to function as an origin of both PeV-energy
neutrinos and HECRs. 

\section{The constraints on the highest-energy cosmic-ray sources}
\label{sec:EeV}

\begin{figure*}
  \includegraphics[width=0.4\textwidth]{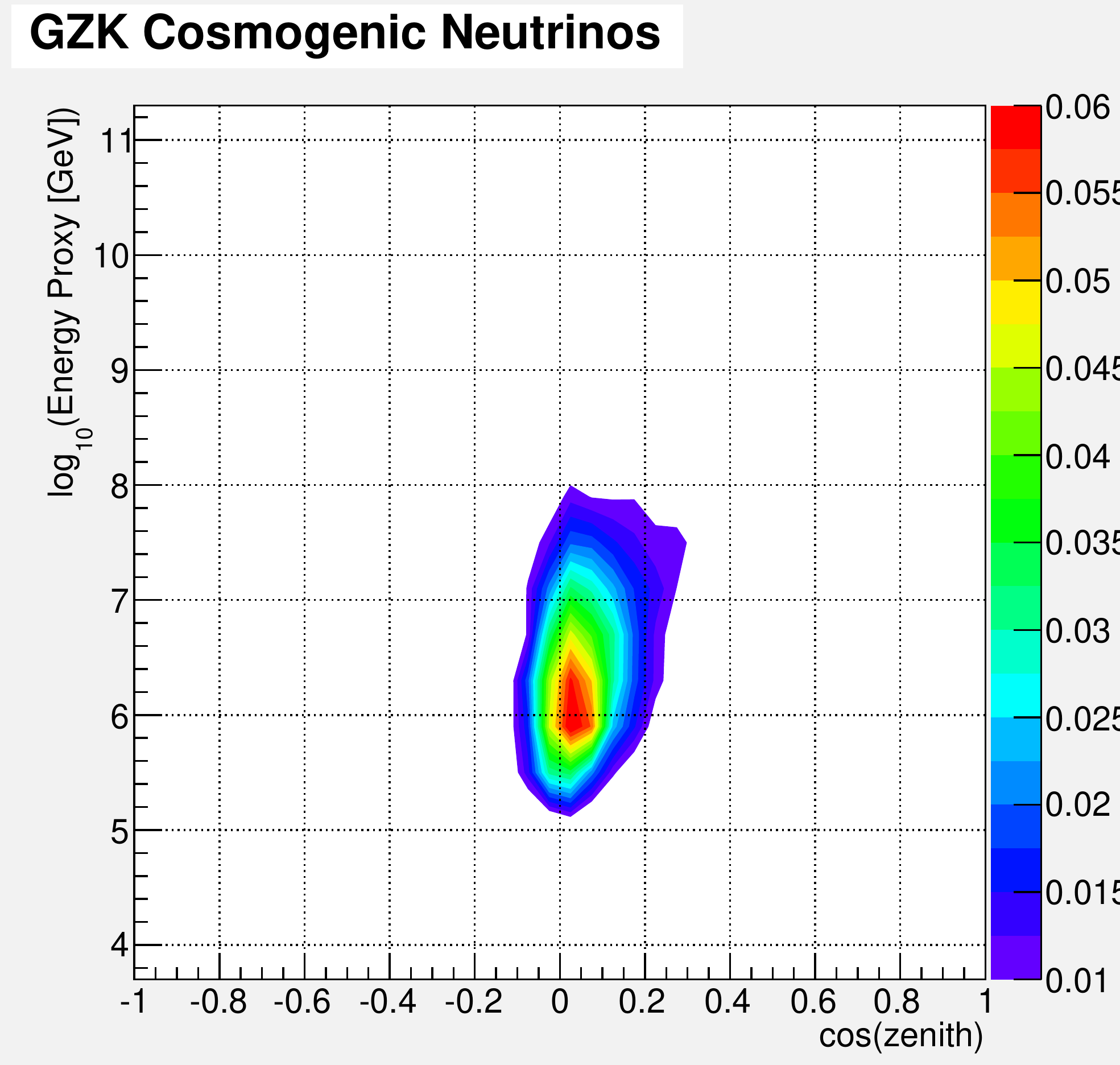}
  \includegraphics[width=0.4\textwidth]{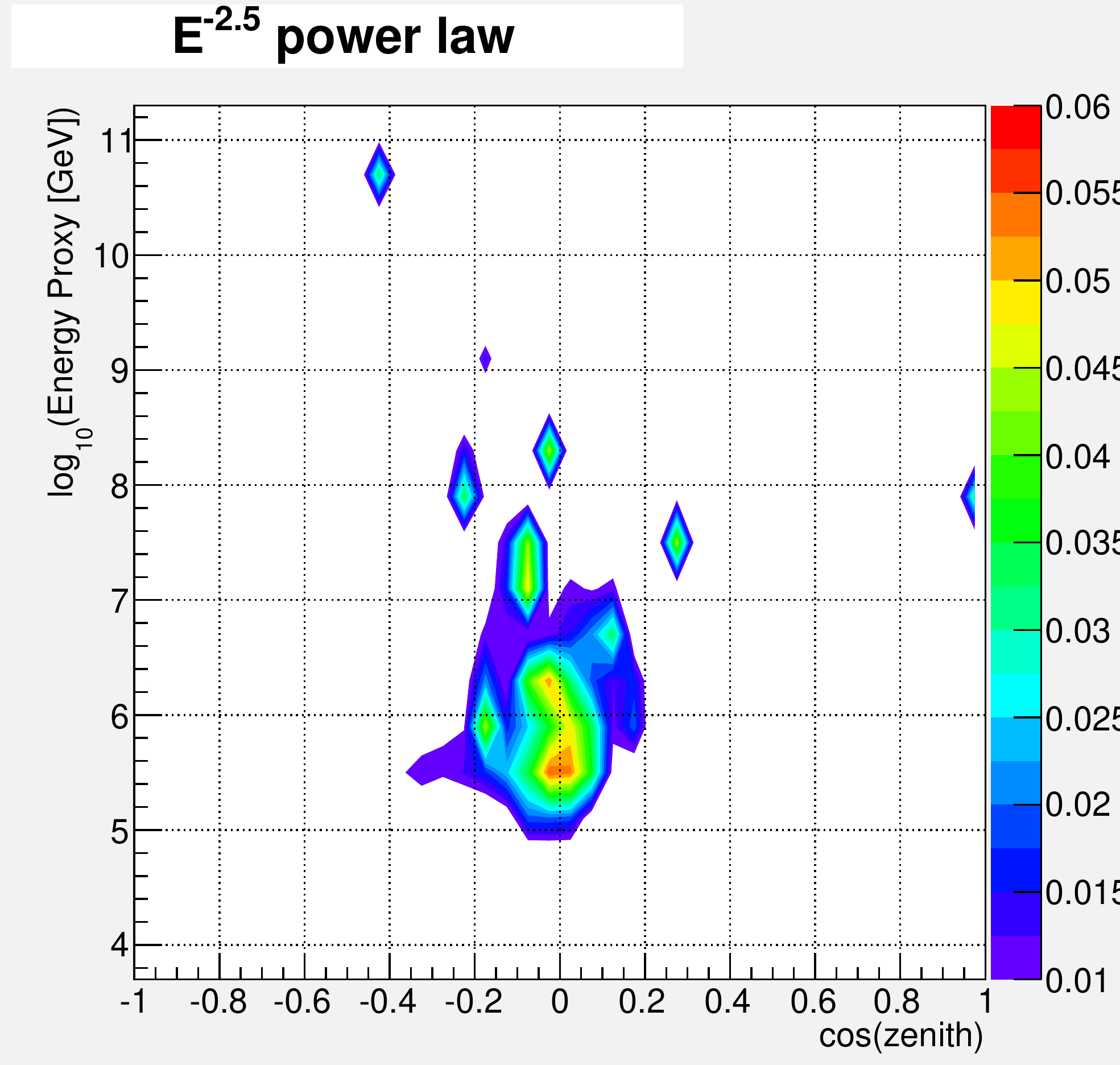}
  \caption{Expected event distributions on the plane of the energy proxy
and cosine of the reconstructed zenith angle seen by IceCube. 
Simulation of the GZK cosmogenic model~\cite{Ahlers2010}
(left panel) and astrophysical neutrino models with a spectrum following $E_\nu^{-2.5}$
are shown with the intensity measured by IceCube~\cite{IceCubeHESE2014}. The $z$ axis displays the number of events seen by the IceCube extremely high-energy analysis
based on the seven year data.}
\label{fig:energy_cosZenith}
\end{figure*}

The analysis of seven years of IceCube data obtained in the search for ultrahigh-energy
neutrinos (energies larger than 10 PeV) has been reported~\cite{EHE2016}.
The analysis was optimized in particular for neutrinos with energies above 100 PeV.
The exposure has reached $\sim10^{17}{\rm cm^2} \sec\ {\rm sr}$ at 1 EeV,
which makes it possible to probe an important region of the parameter space 
in the GZK cosmogenic neutrino models. 
Two events with estimated deposited energies of
2.6 and 0.77 PeV, respectively, were identified in this analysis, but no events were found
in the higher-energy region. This observation presents a serious challenge
to the standard baseline candidates of HECR sources discussed in the literature.

An IceCube simulation was used
to predict the number of events IceCube would detect on the plane of the reconstructed
energy and zenith angle for each model of ultrahigh-energy neutrinos (including GZK cosmogenic neutrinos). 
Figure~\ref{fig:energy_cosZenith} shows examples
from the models. The resolution of the reconstructed deposited energies of energetic
events is less than excellent owing to the stochastic nature of the muon energy loss profile
at PeV energies. Furthermore, IceCube's ability to associate the estimated energy deposit
of an event with its parent neutrino energy is rather limited because only a small fraction
of the neutrino energy is converted to the visible form ({\it i.e.}, the deposited muon energy)
by the IceCube detectors. Nevertheless, Figure~\ref{fig:energy_cosZenith} exhibits
clear differences between two different models. The GZK model yields an event distribution
with an energy peak higher than the softer astrophysical neutrino models.
The events from the GZK model are also distributed more sharply in the horizontal
direction {\it i.e.}, $\cos({\rm zenith})=0$. This is because neutrinos with
energies at the EeV level experience strong absorption effects 
as they propagate through the Earth. These features, as well as
the total event rate (which is equivalent to the normalization of the event distributions
shown in Figure~\ref{fig:energy_cosZenith}), makes it possible to determine which models
are compatible with the observation. The binned Poisson log-likelihood ratio
test was performed. The simulated event distributions 
on the energy--zenith angle plane, such as those shown in Figure~\ref{fig:energy_cosZenith},
give the expected number of events in each bin of the energy proxy and
cosine of the zenith, which were used to construct the binned Poisson likelihood. A log-likelihood ratio was
used as a test statistic, and an ensemble of pseudo-experiments to derive
the test statistical distribution was used to calculate the p-values.

\begin{figure}
  \includegraphics[width=0.4\textwidth]{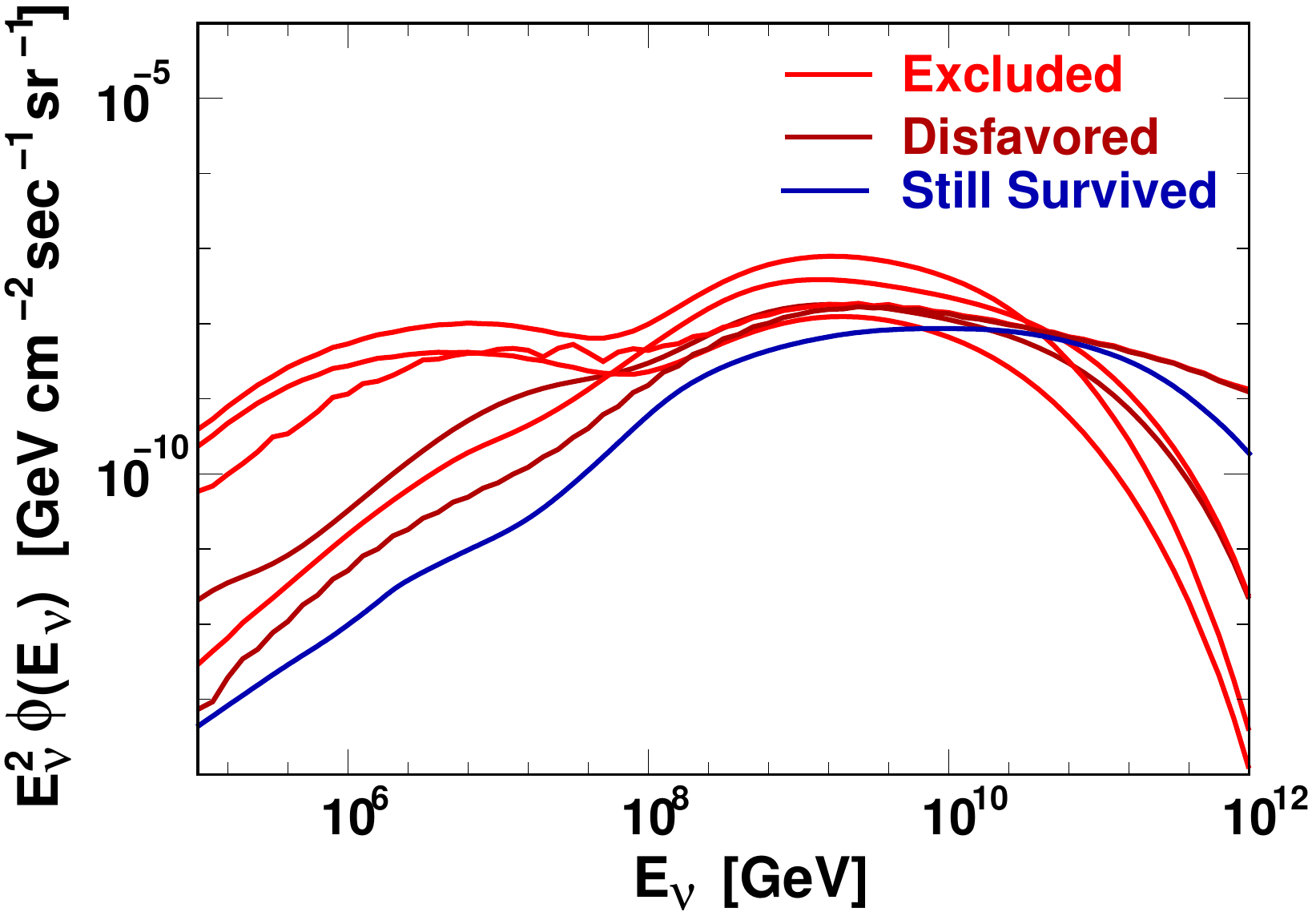}
  \caption{Energy spectra of various cosmogenic neutrino
    models~\cite{Ahlers2010, Kotera2010,Aloisio2015, Yoshida1993}. The spectra shown
by red (brown) curves are rejected (disfavored), with p-values less than 10\% (32\%).
All these models assume a proton-dominated HECR composition.}
\label{fig:gzk_flux}
\end{figure}

The hypothesis that the two observed {\it PeV-ish} events are of GZK cosmogenic origin
is rejected, with a p-value of 0.3\% (which implies that it is incompatible 
with the event distribution shown in the left panel of Figure~\ref{fig:energy_cosZenith})
but is consistent with a generic astrophysical power-law flux such
as $E_\nu^{-2}$ or $E_\nu^{-2.5}$~\cite{EHE2016}. 
This result makes it possible to set an upper limit
on the ultrahigh-energy neutrino flux extending above 10 PeV and thus to also test
the GZK cosmogenic neutrino models using the binned Poisson log-likelihood ratio method.

The various cosmogenic neutrino energy spectra are displayed in
Figure~\ref{fig:gzk_flux}. Many of them are rejected or disfavored by the IceCube
observation. Regardless of where the HECR sources are
and how they accelerate cosmic rays, the emitted HECR protons {\it must} produce
secondary neutrinos by the GZK mechanism as they travel through space.
In this sense, any consequences of these bounds on GZK neutrinos
are considered as robust and model-independent arguments.

We summarize the findings below.

\begin{itemize}
\item Cosmogenic models with the maximal flux allowed
  by the Fermi-LAT measurement~\cite{FermiDiffuse} of 
  the diffuse extragalactic $\gamma-$ray background are rejected.
  This finding implies that the present limits imposed by the neutrino observation are
  at least as stringent as those imposed by the $\gamma-$ray observation.

\item HECR source evolution comparable to the star formation rate (SFR) is beginning to
be constrained. Sources evolving more strongly than the SFR, such as FSRQs and GRBs, are unlikely
to be HECR sources; otherwise, IceCube would have detected cosmogenic-neutrino-induced events already.

\item Any GZK cosmogenic type of energy spectrum must have an intensity below
$E_\nu^2\phi_{\nu_e + \nu_\mu + \nu_\tau}(E_\nu) = 3\times 10^{-9} 
{\rm GeV}\ {\rm cm}^{-2}\ \sec^{-1}\ {\rm sr}^{-1}$ at 100 PeV. This limit rejects the dip transition
model of UHECRs.

\end{itemize}

\begin{figure}
  \includegraphics[width=0.45\textwidth]{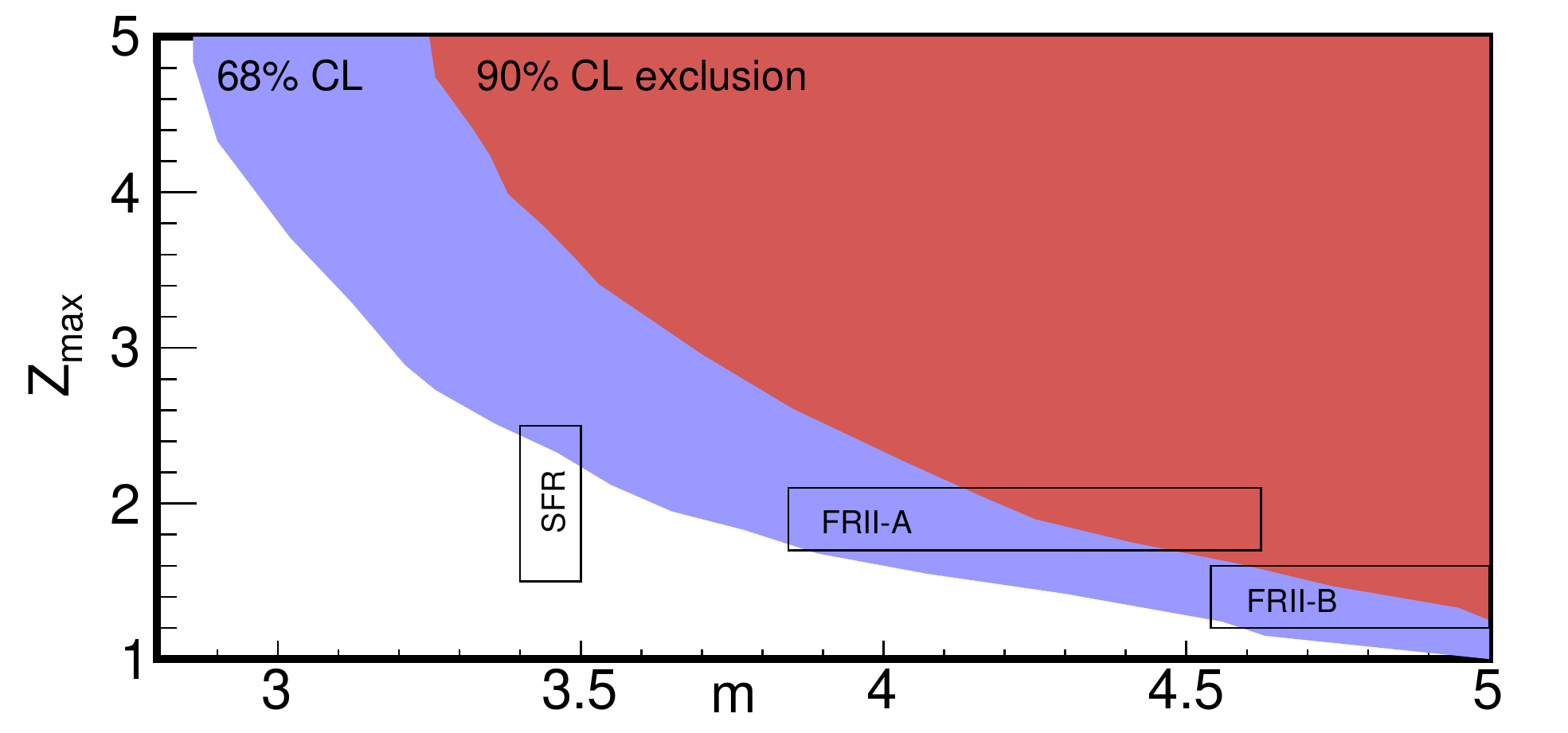}
  \caption{Constraints on HECR source evolution parameters. The emission rate per co-moving volume
    is parameterized as $\psi(z)=(1+z)^m$ with redshifts up to $Z_{\rm max}$. GZK cosmogenic neutrino
    fluxes for various $m$ and $Z_{\rm max}$ values are calculated by the approximated analytical
    formulation~\cite{YoshidaIshihara2012} and used for the likelihood calculation to derive
    the confidence levels. The boxes indicate approximate parameter regions for the SFR~\cite{Beacom} and
    FR-II-A~\cite{FR2-A} and -B~\cite{FR2-B} radio galaxies.}
\label{fig:evolution}
\end{figure}

More generic constraints obtained by the IceCube Collaboration~\cite{EHE2016} by scanning the parameter space for the
source evolution function, $\psi(z)=(1+z)^m$,
extending to the maximal redshift $Z_{\rm max}$ 
are shown in Figure~\ref{fig:evolution}.
The parameterized analytical formula for the cosmogenic fluxes~\cite{YoshidaIshihara2012} is used here.
Because only the CMB is assumed as the target photon field in the parameterization,
the limits are systematically weaker than those on the models that include EBLs.
Approximate regions for the SFR and the evolution
of Fanaroff--Riley type II (FR-II) galaxies are also shown for comparison.
Note that neutrinos yielded at redshifts larger than 2
represent only a minor portion of the total cosmogenic fluxes owing to redshift
dilution~\cite{Kotera2010, YoshidaIshihara2012}.
This is especially true for the cosmogenic neutrino component created
by interactions with the CMB (not the EBL). Considering this fact,
together with the estimation that
the luminosity function of FR-II-type AGNs or FSRQs falls off rapidly at redshifts beyond
$z\simeq 2$, and that the evolution of the SFR becomes more or less constant or falls off
at redshifts beyond $z\sim2.5$, the boxes representing SFR and FR-II evolution
in Figure~\ref{fig:evolution} approximate well their representation by
the generic evolution function $\psi(z)=(1+z)^m$ used in the plot.
One can find that HECR source evolution stronger than the SFR is unlikely.

\begin{figure}
  \includegraphics[width=0.45\textwidth]{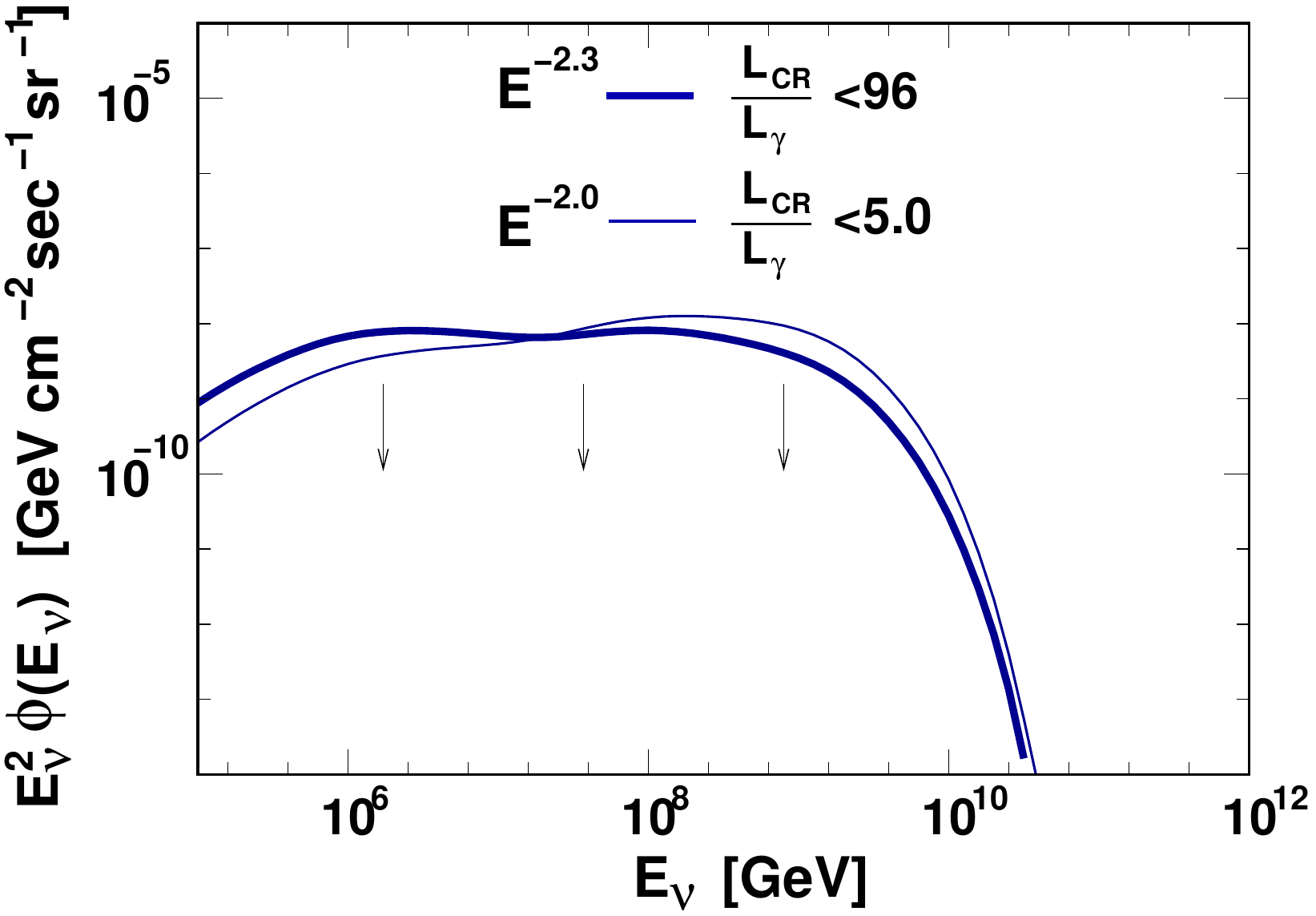}
  \caption{Constraints on the fluxes of astrophysical neutrinos produced
in the inner jets of radio-loud AGNs~\cite{murase2014}. Two bounds for
the UHECR spectral index, $\alpha=2.0$ (thin) and $2.3$ (thick),
are shown.}
\label{fig:agn_nu}
\end{figure}

All the constraints on the HECR origins described so far 
rely on one critical assumption, that HECRs are proton-dominated.
If HECRs are of mixed- or heavy-nuclei composition, the resultant GZK cosmogenic
flux is lower than the proton UHECR case by more than an order of magnitude,
and the present IceCube detection sensitivity cannot reach this low intensity.
It is expected, however, that neutrinos with energies from the PeV level to the EeV level and beyond
may be produced {\it in situ} at the HECR acceleration site. 
The AGN neutrino models are good examples. A recent theoretical study
of ultrahigh-energy neutrino generation in the inner jets 
of radio-loud AGNs~\cite{murase2014} found that, 
taking into account the blazar sequence,
FSRQs can emit PeV--EeV neutrinos, and BL Lac objects
can be HECR (heavy) {\it nuclei} sources~\cite{murase2012}.
The predicted PeV--EeV neutrino intensity is proportional
to the baryon loading factor, that is, the ratio of the UHECR luminosity
to the electromagnetic radiation luminosity $L_{\rm CR}/L_\gamma$.
The null detection of 100 PeV--EeV neutrinos by IceCube thus
bounds this factor.

Figure~\ref{fig:agn_nu} shows the present bound
on the fluxes of neutrinos from radio-loud AGNs
by IceCube~\cite{EHE2016}.
The observed HECR generation rate, $\sim$10 EeV 
($10^{44}\ {\rm erg}\ {\rm Mpc}^{-3}\ {\rm yr}^{-1}$), requires loading factors
of around 3 and 100 for UHECR spectral indices of $\alpha = 2$ and 2.3, respectively.
The present constraints are comparable to or slightly below the values
required for radio-loud AGN inner jets to be responsible for
the majority of UHECRs/HECRs. 
The neutrino observation has started to exclude a sizable parameter
space in the models of AGNs as an origin of HECRs even if
HECRs are composed of heavy nuclei, although this is a model-dependent
argument.

Fast-spinning newborn pulsars are also proposed as candidate sources
of HECRs~\cite{olinto1997}. This proposal predicts a heavy-nuclei-dominated composition
at the highest energies and thus would yield GZK neutrinos too rare
to be detected. However, in this model, the accelerated particles traveling through
the expanding supernova ejecta surrounding the star
produce neutrinos with energies of 100 PeV to $\sim$EeV. The predicted
diffuse neutrino flux from fast pulsars is 
$E_\nu^2 \phi_{\nu_e+\nu_\mu+\nu_\tau}\simeq 
1.1\times 10^{-8} {\rm GeV} {\rm cm^{-2}} \sec^{-1} {\rm sr^{-1}}$,
depending on the source emission evolution, and
is accessible at the IceCube detection sensitivity~\cite{fang2014}.

A binned Poisson log-likelihood test of this model was performed by
the IceCube Collaboration~\cite{EHE2016}. The model is rejected
if the evolution of the source emission history traces the standard SFR, although
it is not ruled out if the emission rate evolves more slowly than the SFR.

\section{Conclusion}
\label{sec:conclusion}

The detection of TeV--PeV neutrinos and the null detection
of EeV neutrinos by IceCube have yielded many insights on the origin of UHECRs, which cannot be probed by other cosmic messengers.
The most popular candidates for UHECR/HECR sources, GRBs and radio-loud AGNs,
have now faced serious challenges from recent results in neutrino astronomy.
GRBs and AGNs can still contribute to the observed bulk
of the highest-energy cosmic rays but are unlikely to be their {\it dominant} sources.
Astronomical objects tracing the standard SFR or evolving much more slowly
are needed to explain the observation without fine-tuning.
If the highest-energy cosmic rays are not proton-dominated,
these constraints are certainly relaxed. The model-dependent tests
described here, however, have already placed limits on some of
the parameter space of the AGN/pulsar scenarios.

There is a loophole: a hypothesis that any high-energy neutrino emission
does not involve cosmic-ray emission. A good example is the GRB choked jet
model~\cite{chokedGRB}. In dense environments, the optical depth $\tau_0\gg 1$,
which implies that all the proton專 energy is
converted into neutrinos; {\it i.e.}, the observed UHECRs and
neutrinos are not directly connected.

How can we identify UHECR/HECR sources that evolve
at the usual SFR, or even more slowly?
Real-time multi-messenger observation
triggered by high-energy neutrinos is a possible answer. IceCube has launched
the Gamma-ray Coordinates Network-based alert delivery system~\cite{icecubeGCN}.
The search algorithms for Extremely High-Energy  neutrinos~\cite{icecubePeV2013, EHE2016}
and High-Energy Starting Events~\cite{icecubeHESE2013, IceCubeHESE2014},
the analysis channels that discovered the high-energy cosmic neutrinos, are now running
in real time at IceCube's South Pole data servers.
Once a high-energy-neutrino-induced event is detected, an alert is sent immediately
to trigger follow-up observations by other astronomical instruments.
If UHECR/HECR sources are transient neutrino sources,
we may be able to identify them by follow-up detection with optical/X-ray/$\gamma$-ray/radio
telescopes. This is probably a promising way to approach identification of
the yet-unknown origins of high-energy cosmic rays.

\section*{Acknowledgments}
I am grateful to the CRIS 2016 organizers for
their warm hospitality.
I acknowledge my colleagues in the IceCube Collaboration
for useful discussions and suggestions.
I also appreciate the input of Kunihito Ioka, Kumiko Kotera, Kohta Murase, 
and Hajime Takami on the theoretical arguments.
Special thanks go to Aya Ishihara, who has worked together
on the analyses of extremely high-energy neutrinos 
for many years. This work is supported by JSPS
Grants-in-Aid for Scientific Research (Project \#25105005 and \#25220706).




\nocite{*}
\bibliographystyle{elsarticle-num}
\bibliography{martin}

\begin{thebibliography}{00}


\bibitem{icecubePeV2013}
M.~G.~Aartsen {\it et al.} (IceCube Collaboration), Phys.~Rev.~Lett. {\bf 111}, 021103 (2013).
\bibitem{icecubeHESE2013}
  M.~G.~Aartsen {\it et al.} (IceCube Collaboration), Science {\bf 342}, 1242856 (2013).
\bibitem{GZK}
  K.~Greisen, Phys. Rev. Lett. {\bf 16}, 748 (1966);
  G. T. Zatsepin and V. A. Kuzmin, Pisma Zh. Eksp. Teor. Fiz. {\bf 4}, 114 (1966)
  [JETP Lett. {\bf 4}, 78 (1966)].
\bibitem{BZ}
V. S. Berezinsky and G. T. Zatsepin, Phys. Lett. {\bf 28B}, 423 (1969).  
\bibitem{YoshidaTakami2014}
S. Yoshida and H. Takami, Phys. Rev.~D {\bf 90}, 123012 (2014).
\bibitem{MESE}
M.~G.~Aartsen {\it et al.} (IceCube Collaboration), Phys.~Rev.~D {\bf 91}, 022001 (2015).
\bibitem{diffuse_nu}
M.~G.~Aartsen {\it et al.} (IceCube Collaboration), Phys.~Rev.~Lett. {\bf 115}, 081102 (2015).
\bibitem{icetop2013}
M.~G.~Aartsen {\it et al.} (IceCube Collaboration), Phys.~Rev.~D {\bf 88}, 042004 (2013).
\bibitem{auger_density}
P.~Abreu {\it et al.} (Pierre Auger Collaboration), J.~Cosmol.~Astropart.~Phys. {\bf 05}, 009 (2013).
\bibitem{dip_model}
  V.~Berezinsky, A.~Gazizov, and S.~Grigorieva, Phys.~Rev.~D {\bf 74}, 043005 (2006).
\bibitem{EHE2016}
  M.~G.~Aartsen {\it et al.} (IceCube Collaboration), Phys.~Rev.~Lett. {\bf 117}, 241101 (2016).
\bibitem{Ahlers2010}
M.~Ahlers {\it et al.}, Astropart. Phys. {\bf 34}, 106 (2010).
\bibitem{IceCubeHESE2014}
M.~G.~Aartsen {\it et al.} (IceCube Collaboration), Phys.~Rev.~Lett. {\bf 113}, 101101 (2014).
\bibitem{Kotera2010}
  K.~Kotera, D.~Allard, and A.~Olinto, J.~Cosmol.~Astropart.~Phys. {\bf 2010}, 013 (2010).
\bibitem{Aloisio2015}
  R. Aloisio {\it et al.}, J.~Cosmol.~Astropart.~Phys. {\bf 10}, 006 (2015).
\bibitem{FermiDiffuse}
M.~Ackermann {\it et al.} (Fermi-LAT Collaboration), Astrophys.~J. {\bf 799}, 86 (2015).
\bibitem{Yoshida1993}
S.~Yoshida and M.~Teshima, Prog.~Theor.~Phys. {\bf 89}, 833 (1993).
\bibitem{YoshidaIshihara2012}
S.~Yoshida and A.~Ishihara, Phys.~Rev.~D {\bf 85}, 063002 (2012).
\bibitem{Beacom}
  A.~M.~Hopkins and J.~F.~Beacom, Astrophys.~J. {\bf 651}, 142 (2006).
\bibitem{FR2-A}
  Y.~Inoue and T.~Totani, Astrophys.~J. {\bf 702}, 523 (2009).
\bibitem{FR2-B}
  M. Ajello {\it et al.}, Astrophys.~J. {\bf 751}, 108 (2012).
\bibitem{murase2014}
K.~Murase, Y.~Inoue, and C.~D.~Dermer, Phys. Rev. D {\bf 90}, 023007 (2014).
\bibitem{murase2012}
K.~Murase, C.~D.~Dermer, H.~Takami, and G.~Migliori, Astrophys.~J. {\bf 749}, 63 (2012).
\bibitem{olinto1997}
A.~Venkatesan, M.~C.~Miller, and A.~V.~Olinto, Astrophys.~J. {\bf 484}, 323 (1997).
\bibitem{fang2014}
Ke.~Fang, K.~Kotera, K.~Murase, and A.~V.~Olinto, Phys. Rev. D {\bf 90}, 103005 (2014).
\bibitem{chokedGRB}
  P.~M市z㎎os and E.~Waxman, Phys. Rev. Lett. {\bf 87}, 171102 (2001);
  K.~Murase and K.~Ioka, Phys. Rev. Lett. {\bf 111}, 121102 (2013).
\bibitem{icecubeGCN}
http://gcn.gsfc.nasa.gov/amon.html
  
\end{thebibliography}



\end{document}